\documentclass[twoside]{article}
\usepackage{fleqn,espcrc2}

\usepackage{graphicx,psfrag}
\usepackage[figuresright]{rotating}

\newcommand{\AmS}{{\protect\the\textfont2
  A\kern-.1667em\lower.5ex\hbox{M}\kern-.125emS}}

\title{The Perfect Laplace Operator for Non-Trivial Boundaries}

\author{S. Hauswirth\address{Institute of Theoretical
        Physics, University of Bern\\Sidlerstrasse 5, CH-3012 Bern,
        Switzerland}}
       
\begin{document}

\begin{abstract}
The application of Renormalization Group (RG) methods to find perfect
discretizations of partial differential 
equations is a promising but little investigated approach.
We calculate the classically perfect fixed-point Laplace
operator for boundaries of non-trivial shape analytically and numerically
and present a parametrization that can be used for solving the Poisson
equation. 
\vspace{1pc}
\end{abstract}

\maketitle

\section{INTRODUCTION}

In Lattice QCD, the perfect action provides a discrete formulation of the
theory 
which is free from lattice artifacts \cite{hasenfratz_niedermayer:94}. Few
attempts have so far been made to examine
perfect discretizations also for classical problems like the numerical
solution of differential equations \cite{katz_wiese:97,goldenfeld:98,hou:00}.
We present an exploratory study of a perfect differential operator in presence
of 
non-trivial boundary condiations, as they appear in classical field theory
applications. This work is in some more detail also presented in \cite{hauswirth:00}.

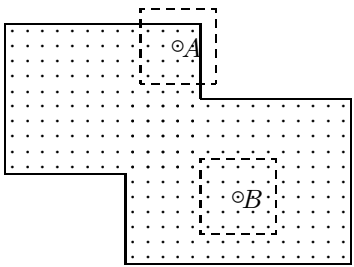
\begin{figure}[h]
\setlength{\unitlength}{0.1cm}
\begin{center}
\begin{picture}(50,17)(0,0)
  \put(18,12){\dashbox(10,10)}
  \put(23.5,15.5){$A$}
  \put(23,17){\circle{1.5}}
  \put(31.5,-4.5){$B$}
  \put(26,-8){\dashbox(10,10)}
  \put(31,-3){\circle{1.5}}
  \put(0,0){\line(1,0){16}}
  \put(0,0){\line(0,1){20}}
  \put(0,20){\line(1,0){26}}
  \put(26,20){\line(0,-1){10}}
  \put(26,10){\line(1,0){20}}
  \put(16,0){\line(0,-1){12}}
  \put(16,-12){\line(1,0){30}}
  \put(46,10){\line(0,-1){22}}
  \multiput(1,1)(0,2){10}{\multiput(0,0)(2,0){13}{\circle*{0.4}}}
  \multiput(17,-11)(0,2){11}{\multiput(0,0)(2,0){15}{\circle*{0.4}}}
\end{picture}
\end{center}
\caption{A two-dimensional bounded region.}\label{fig_region}
\end{figure}

Our toy model is the Laplace
equation on a two-dimensional region with fixed boundaries which are built
from vertical and horizontal lines as shown in Fig.~\ref{fig_region}.

Solving such a problem numerically with an improved discretization of the
Laplace operator reduces the need for high resolution of the underlying
mesh. The most improved discretization -- the classically perfect fixed point
operator -- is completely free of discretization errors. 
Perfect operators have non-zero couplings spreading over an infinite volume,
so for practical use they have to be truncated. Suppose we want to use an
operator with only nearest- and
next-to-nearest neighbour couplings. Then we have to provide a set of
operators which differ in the position of the center point with respect to the
boundary (e.g. for point $A$ in Fig.~\ref{fig_region} we need another operator
than 
for point $B$) as symbolized by the icons in Fig.~\ref{fig_icons}:

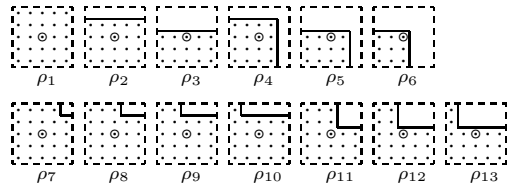
\begin{figure}[h]
\vspace{-5mm}
\setlength{\unitlength}{0.08cm}
\begin{center}
\begin{picture}(140,10)(-5,12)
  \put(0,16){\dashbox(10,10)}
  \put(5,21){\circle{1.5}}
  \multiput(1,17)(0,2){5}{\multiput(0,0)(2,0){5}{\circle*{0.4}}}
  \put(4,13.5){\scriptsize$\rho_1$}

  \put(12,16){\dashbox(10,10)}
  \put(12,24){\line(1,0){10}}
  \put(17,21){\circle{1.5}}
  \multiput(13,17)(0,2){4}{\multiput(0,0)(2,0){5}{\circle*{0.4}}}
  \put(16,13.5){\scriptsize$\rho_{2}$}

  \put(24,16){\dashbox(10,10)}
  \put(24,22){\line(1,0){10}}
  \put(29,21){\circle{1.5}}
  \multiput(25,17)(0,2){3}{\multiput(0,0)(2,0){5}{\circle*{0.4}}}
  \put(28,13.5){\scriptsize$\rho_{3}$}

  \put(36,16){\dashbox(10,10)}
  \put(36,24){\line(1,0){8}}
  \put(44,16){\line(0,1){8}}
  \put(41,21){\circle{1.5}}
  \multiput(37,17)(0,2){4}{\multiput(0,0)(2,0){4}{\circle*{0.4}}}
  \put(40,13.5){\scriptsize$\rho_{4}$}

  \put(48,16){\dashbox(10,10)}
  \put(48,22){\line(1,0){8}}
  \put(56,16){\line(0,1){6}}
  \put(53,21){\circle{1.5}}
  \multiput(49,17)(0,2){3}{\multiput(0,0)(2,0){4}{\circle*{0.4}}}
  \put(52,13.5){\scriptsize$\rho_{5}$}

  \put(60,16){\dashbox(10,10)}
  \put(60,22){\line(1,0){6}}
  \put(66,16){\line(0,1){6}}
  \put(65,21){\circle{1.5}}
  \multiput(61,17)(0,2){3}{\multiput(0,0)(2,0){3}{\circle*{0.4}}}
  \put(64,13.5){\scriptsize$\rho_{6}$}

  \put(0,0){\dashbox(10,10)}
  \put(8,8){\line(1,0){2}}
  \put(8,8){\line(0,1){2}}
  \put(5,5){\circle{1.5}}
  \multiput(1,1)(0,2){4}{\multiput(0,0)(2,0){5}{\circle*{0.4}}}
  \multiput(1,9)(0,2){1}{\multiput(0,0)(2,0){4}{\circle*{0.4}}}
  \put(4,-2.5){\scriptsize$\rho_{7}$}

  \put(12,0){\dashbox(10,10)}
  \put(18,8){\line(1,0){4}}
  \put(18,8){\line(0,1){2}}
  \put(17,5){\circle{1.5}}
  \multiput(13,1)(0,2){4}{\multiput(0,0)(2,0){5}{\circle*{0.4}}}
  \multiput(13,9)(0,2){1}{\multiput(0,0)(2,0){3}{\circle*{0.4}}}
  \put(16,-2.5){\scriptsize$\rho_{8}$}

  \put(24,0){\dashbox(10,10)}
  \put(28,8){\line(1,0){6}}
  \put(28,8){\line(0,1){2}}
  \put(29,5){\circle{1.5}}
  \multiput(25,1)(0,2){4}{\multiput(0,0)(2,0){5}{\circle*{0.4}}}
  \multiput(25,9)(0,2){1}{\multiput(0,0)(2,0){2}{\circle*{0.4}}}
  \put(28,-2.5){\scriptsize$\rho_{9}$}

  \put(36,0){\dashbox(10,10)}
  \put(38,8){\line(1,0){8}}
  \put(38,8){\line(0,1){2}}
  \put(41,5){\circle{1.5}}
  \multiput(37,1)(0,2){4}{\multiput(0,0)(2,0){5}{\circle*{0.4}}}
  \multiput(37,9)(0,2){1}{\multiput(0,0)(2,0){1}{\circle*{0.4}}}
  \put(40,-2.5){\scriptsize$\rho_{10}$}

  \put(48,0){\dashbox(10,10)}
  \put(54,6){\line(1,0){4}}
  \put(54,6){\line(0,1){4}}
  \put(53,5){\circle{1.5}}
  \multiput(49,1)(0,2){3}{\multiput(0,0)(2,0){5}{\circle*{0.4}}}
  \multiput(49,7)(0,2){2}{\multiput(0,0)(2,0){3}{\circle*{0.4}}}
  \put(52,-2.5){\scriptsize$\rho_{11}$}

  \put(60,0){\dashbox(10,10)}
  \put(64,6){\line(1,0){6}}
  \put(64,6){\line(0,1){4}}
  \put(65,5){\circle{1.5}}
  \multiput(61,1)(0,2){3}{\multiput(0,0)(2,0){5}{\circle*{0.4}}}
  \multiput(61,7)(0,2){2}{\multiput(0,0)(2,0){2}{\circle*{0.4}}}
  \put(64,-2.5){\scriptsize$\rho_{12}$}

  \put(72,0){\dashbox(10,10)}
  \put(74,6){\line(1,0){8}}
  \put(74,6){\line(0,1){4}}
  \put(77,5){\circle{1.5}}
  \multiput(73,1)(0,2){3}{\multiput(0,0)(2,0){5}{\circle*{0.4}}}
  \multiput(73,7)(0,2){2}{\multiput(0,0)(2,0){1}{\circle*{0.4}}}
  \put(76,-2.5){\scriptsize$\rho_{13}$}

\end{picture}
\end{center}
\caption{The set of Laplace operators $\rho_i(r_1,r_2)$ near straight
  boundaries, including couplings with $r_i\leq 2$ ($i=1,2$).}\label{fig_icons}
\end{figure}

We construct such a set of operators with the perfect action
formalism. In order to check how well this construction works, we
examine how the boundary influences the perfect Laplacian. We find that not
only the operator itself is local (i.e. its couplings decrease 
exponentially with distance), but also the effect of the boundary on the
operator (so that for example in point $B$ we don't have to care about any
boundaries).

\section{THE PERFECT LAPLACIAN}

 The perfect
Laplace operator in $d$ dimensions can be 
calculated from the fixed point action of a free real scalar field with
the continuum action
\begin{equation} \label{contfreeaction}
{\mathcal{A}}^{cont.}(\phi) = \frac{1}{2}\int\! d^dx\ 
\partial_\mu\phi(x)\partial_\mu\phi(x).
\end{equation}
The corresponding equation of motion is the Laplace equation. A general
discretization of the action (\ref{contfreeaction}) contains terms which couple
the field at one lattice site to the field at another one:
\begin{equation} \label{quadratic_form}
{\mathcal{A}}(\phi) = \frac{1}{2} \sum_{n,r}\phi_n \rho(r) \phi_{n+r},
\end{equation}
with the couplings $\rho(r)$, $r=(r_1,\dots,r_d)$. For the standard Laplacian in two
dimensions, $\rho(0)=4$ and $\rho(r)=-1\ (\forall |r|=1)$. 
A RG transformation of the form
\begin{equation} \label{generalrgt}
  C \cdot e^{ -{\mathcal{A}}^\prime(\chi)} = \prod_n\int\! d\phi_n \
  e^{-{\mathcal{A}}(\phi) - {\mathcal{T}}(\chi,\phi)} .
\end{equation}
relates the actions $\mathcal{A}(\phi)$ on a fine and $\mathcal{A}^\prime(\chi)$ on a coarse lattice with lattice units $a$ and
$2a$, respectively. $C$ is a normalization constant and $\mathcal{T}$ the
blocking kernel
\begin{equation} \label{gauss_block_kernel}
{\mathcal{T}}(\chi,\phi) = 2\kappa \sum_{n_B}(\chi_{n_B} -
  b\cdot\frac{1}{2^d}\sum_{n\in n_B} \phi_n)^2.
\end{equation}
where $\kappa$ is a parameter used for optimizing the fixed point Laplacian
in terms of locality. We use the kernel (\ref{gauss_block_kernel}) with $\kappa=2$ in the following, as this gives a very local
operator (the couplings decrease exponentially $\propto e^{-\gamma r}$
with a large decay coefficient $\gamma\approx 3.5$) \cite{ruefenacht:98,bietenholz:99}.
The fixed point of the RG transformation (\ref{generalrgt})
can be calculated analytically
\cite{bell_wilson:74}. In momentum space, the result is
$$
\frac{1}{\tilde\rho^*(q)} = \sum_{l\in {\mathbf Z}^d} \frac{1}{(q+2\pi l)^2}
\prod_\mu \frac{\sin^2(\frac{q_\mu}{2}+\pi l_\mu)}{(\frac{q_\mu}{2}+\pi
  l_\mu)^2} + \frac{1}{3\kappa}. 
$$
The couplings $\rho^*(n)$ of the fixed point Laplacian are found by Fourier
transformation. From this result we derive the element $\rho_1$ of the set of
operators in Fig.~\ref{fig_icons}.

\section{SQUARE-SHAPED BOUNDARIES}

\begin{figure}[h]
\vspace{-5mm}
\setlength{\unitlength}{0.1cm}
\begin{center}
\begin{picture}(30,30)(-5,-5)
\multiput(0,5)(0,10){2}{\line(1,0){20} }
\multiput(5,0)(10,0){2}{\line(0,1){20} }
\multiput(5,-5)(10,0){2}{\dashbox{0.5}(0,30) }
\multiput(-5,5)(0,10){2}{\dashbox{0.5}(30,0) }
\put(0,0){\makebox(0,0){$\phi$}}
\put(10,0){\makebox(0,0){$-\phi$}}
\put(20,0){\makebox(0,0){$\phi$}}
\put(0,10){\makebox(0,0){$-\phi$}}
\put(10,10){\makebox(0,0){$\phi$}}
\put(20,10){\makebox(0,0){$-\phi$}}
\put(0,20){\makebox(0,0){$\phi$}}
\put(10,20){\makebox(0,0){$-\phi$}}
\put(20,20){\makebox(0,0){$\phi$}}
\end{picture}
\end{center}
\vspace{-1cm}
\caption{The field is periodically mirrored on the walls of
  squares.}\label{fig_mirror}
\end{figure}
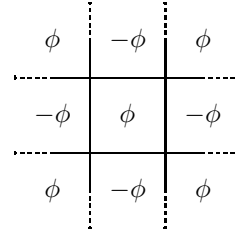

Consider a square region with $\phi=0$ on the boundaries and side length $Na$.
In this case, the perfect Laplacian can be calculated analytically.
The easiest way to construct it in $d=2$ is from symmetries: Extend the
field $\phi$ onto the 
whole plane by periodically mirroring it at the boundary
with alternating sign
\begin{equation}\label{wrap_field}
  \phi(x_1,x_2) = (-1)^{k_1+k_2}
  \phi\bigl((-1)^{k}x+2lN\bigr),
\end{equation}
for any $k_i=0,1$ and $l_i\in{\mathbf Z}$, ($i=1,2$). In Fig.~\ref{fig_mirror}, the
square in the center shows the region we are interested in, the field $\phi$ is
mirrored at all straight lines and its sign is given explicitly.

In the perfect Laplace equation for the whole plane (with the perfect
Laplacian $\rho^*(n)$), the sum over all
lattice points splits up into sums over $N$-squares
\begin{equation}
  \sum_r\rho^*(r)\phi_{n+r} = \sum_{l_i\in{\mathbf Z}}
  \sum_{k_i=0}^1 \sum_{s_i=0}^{N-1} \rho^*(r-n)\phi_r = 0,
\end{equation}
where the variable $r$ running over the lattice points is given by
$r_i = (-1)^{k_i} s_i-k_i+2l_iN$
for $i=1,2$. Using the symmetry $\rho^*(-r_1,r_2)=\rho^*(r_1,r_2)$ and the
relation (\ref{wrap_field}), 
the perfect Laplacian for square boundaries is given by the infinite sum of 
couplings 
$$
\rho^*(n,r) = \sum_{l_i\in{\mathbf Z}} \sum_{k_i=0}^1
(-1)^{k_1+k_2} 
\rho^*\bigl(r+2kn+k+2lN\bigr), 
$$
with both $n$ and $n+r$ lying inside the boundary. The sum converges very fast
because of the locality of $\rho^*(r)$. From this result we derive the
operators $\rho_2$--$\rho_6$.

The perfect Laplace operator for square boundaries can also be recieved by
performing the RG transformation 
analytically in Fourier space, using basis functions $\Psi_{Q}$ which enforce
the condition $\phi=0$ on the 
boundary. The result is \cite{hauswirth:00}
$$
  \frac{1}{\rho^{*}(Q)} = \sum_{l=-\infty}^{\infty} \frac{1}{(Q+2\pi l)^2}
  \prod_{i=1}^{2}\frac{\sin^2\frac{Q_i}{2}}{(\frac{Q_i}{2}+\pi l_i)^2}
  + \frac{1}{3\kappa}.
$$
The couplings of the perfect Laplacian are found by reverse Fourier
transformation $
  \rho^{*}(n,n^{\prime}) =
  1/N^2 \sum_Q \Psi_Q(n) \Psi_Q(n^{\prime}) \cdot \rho^{*}(Q)$
with $\Psi_Q(n) = \xi_{Q_1}(n_1)\cdot \xi_{Q_2}(n_2)$ and
\begin{equation} \label{fq}
  \xi_{Q_i}(n_i) = \left\{
    \begin{array}{ll}
    \sqrt{2}\sin (Q_i(n_i+\frac{1}{2})) &,Q_i\ne\pi, \\
    \sin (Q_i(n_i+\frac{1}{2})) &,Q_i=\pi.
    \end{array} \right.
\end{equation}
($Q_i=k_i\pi/N$,
$k_i=1,\dots,N$.)
A comparison of the results for the two different constructions shows that
they provide the same couplings. They are also in agreement with
\cite{katz_wiese:97}.

\begin{figure}
\vspace{-5mm}
\begin{center}
\psfrag{n1}{$n^\prime_1$}
\psfrag{n2}{\hspace{-0.5cm}\vspace{-0.1cm} $n^\prime_2$}
\psfrag{rho_label}{\hspace{1cm}$\rho^*$} 
\psfrag{0}{\scriptsize\hspace{-2mm}\vspace{-0.5mm} $0$}
\psfrag{5}{\scriptsize\hspace{-1mm}\vspace{-0.5mm}$5$}
\psfrag{10}{\scriptsize\hspace{-0.5mm}\vspace{-0.5mm}$10$}
\psfrag{15}{\scriptsize\hspace{-0.5mm}\vspace{-0.5mm}$15$}
\psfrag{20}{\scriptsize\hspace{-1.5mm} $20$}
\psfrag{-20}{\scriptsize\hspace{-2mm} $-20$}
\hspace{-1cm}\includegraphics[width=8cm]{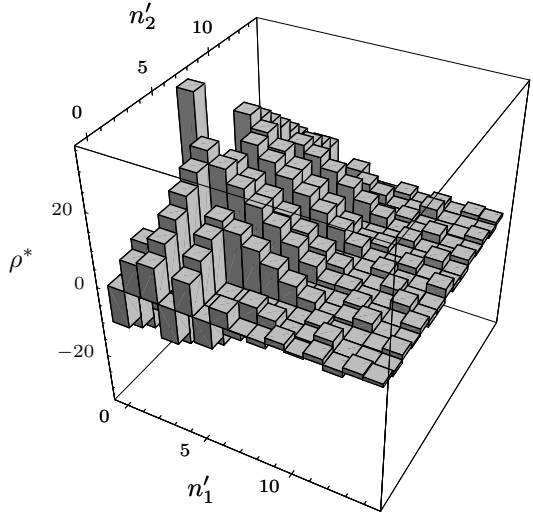}
\end{center}
\vspace{-20mm}
\caption{The couplings $\rho^*(n^\prime_1,n^\prime_2)$ of the perfect Laplacian near a
  wall on a logarithmic scale with preserved sign.} \label{fig_wall}
\end{figure}

\section{CONCAVE CORNERS}

In the case where the boundaries form a concave corner (as in $\rho_7$--$\rho_{13}$), the perfect operator
can't be calculated analytically. Therefore we perform a numerical RG
transformation iteratively on a $N=32$ lattice with L-shaped boundaries as
shown in Fig.~\ref{fig_lshape}:

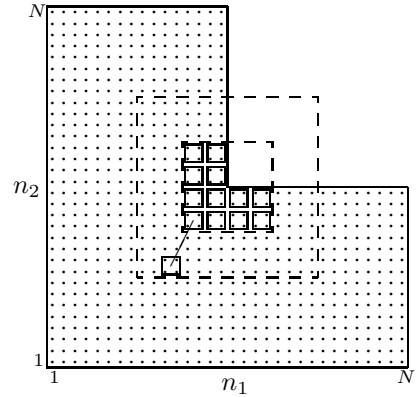
\begin{figure}[h]
\vspace{-10mm}
\setlength{\unitlength}{1.5mm}
\begin{center}
\begin{picture}(32,30)(0,6)
\put(0,0){\line(0,1){32}}
\put(0,0){\line(1,0){32}}
\put(32,0){\line(0,1){16}}
\put(0,32){\line(1,0){16}}
\put(16,16){\line(0,1){16}}
\put(16,16){\line(1,0){16}}
\put(8,8){\dashbox(16,16)}
\put(12,12){\dashbox(8,8)}
\put(10.275,8.275){\framebox(1.45,1.45)}
\put(11,9){\line(1,2){2}}
\multiput(12.275,12.275)(2,0){4}{\multiput(0,0)(0,2){2}{\framebox(1.45,1.45)}}
\multiput(12.275,16.275)(2,0){2}{\multiput(0,0)(0,2){2}{\framebox(1.45,1.45)}} 
\multiput(0.5,0.5)(1,0){32}{\multiput(0,0)(0,1){16}{\circle*{0.15}}}
\multiput(0.5,16.5)(1,0){16}{\multiput(0,0)(0,1){16}{\circle*{0.15}}}
\put(15.5,-2){$n_1$}
\put(-3,15.5){$n_2$}
\put(-1.2,0.1){\scriptsize $1$}
\put(-1.7,31.1){\scriptsize $N$}
\put(0.2,-1.4){\scriptsize $1$}
\put(31,-1.4){\scriptsize $N$}
\end{picture}
\end{center}
\caption{The lattice used to calculate the perfect Laplacian near a concave
  corner. Points denote the fine and boxes the coarse lattice sites.}
  \label{fig_lshape} 
\end{figure}

For quadratic actions and blocking kernels,
the RG transformation (\ref{generalrgt})
can be written as a minimizing condition for the fine field $\phi$:
\begin{equation}\label{eq_minstep}
{\mathcal{A}}^\prime(\chi) = \min_\phi \left[{\mathcal{A}}(\phi) +
  {\mathcal{T}}(\chi,\phi) \right] + \mbox{const.}
\end{equation}
To get close to a fixed point, we have to iterate this RGT step. The
results of the previous step -- which are the couplings of the resulting
coarse
action -- are then used as an input for the next step, that is as a new
starting guess for the fine action ${\mathcal{A}}(\phi)$.
After ${\mathcal{O}}(20)$ iterations, we find a very close
approximation to the couplings of the fixed point action. 

We look for the couplings of the coarse field inside
the large dashed box. Outside the small dashed box, we
use in every
RGT step the operators $\rho_1$--$\rho_6$ for the fine action. For the couplings inside the small dashed box, we use the standard
Laplacian in the first iteration, and afterwards the result of the previous
iteration. As a result we get the operators $\rho_7$--$\rho_{13}$, completing
our set.

\section{RESULTS}

\begin{figure}[h]
\vspace{-5mm}
\begin{center}
\psfrag{xlabel}[][]{\hspace{3mm}\vspace{-2mm}$n_1$}
\psfrag{ylabel}[][]{\hspace{0cm}\vspace{0.6cm}$\Delta\rho^*$}
\psfrag{-30}{\hspace{-2mm}\scriptsize$-30$}
\psfrag{-25}{\hspace{-2mm}\scriptsize$-25$}
\psfrag{-20}{\hspace{-2mm}\scriptsize$-20$}
\psfrag{-15}{\hspace{-2mm}\scriptsize$-15$}
\psfrag{-10}{\hspace{-2mm}\scriptsize$-10$}
\psfrag{-5}{\hspace{-2mm}\scriptsize$-5$}
\psfrag{0}{\vspace{-0.5mm}\scriptsize$0$}
\psfrag{1}{\vspace{-0.5mm}\scriptsize$1$}
\psfrag{2}{\vspace{-0.5mm}\scriptsize$2$}
\psfrag{3}{\vspace{-0.5mm}\scriptsize$3$}
\psfrag{4}{\vspace{-0.5mm}\scriptsize$4$}
\psfrag{5}{\vspace{-0.5mm}\scriptsize$5$}
\psfrag{6}{\vspace{-0.5mm}\scriptsize$6$}
\psfrag{7}{\vspace{-0.5mm}\scriptsize$7$}
\psfrag{8}{\vspace{-0.5mm}\scriptsize$8$}
\psfrag{(0,1)-couplinggg}{\scriptsize $r=(0,1)$}
\psfrag{(1,0)-coupling}{\scriptsize $r=(1,0)$}
\psfrag{(1,1)-coupling}{\scriptsize $r=(1,1)$}
\includegraphics[width=7cm]{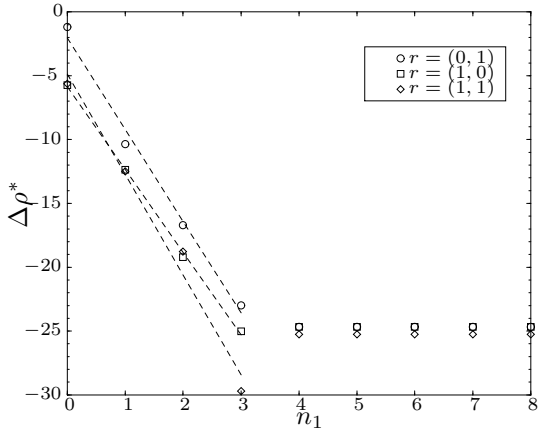}
\end{center}
\vspace{-10mm}
\caption{The relative difference $\Delta\rho^* = \log\left|
    \bigl( \rho^*(n,r)-\rho^*(r) \bigr) / \rho^*(r)\right|$ between the
  couplings near a 
  wall and the respective couplings of the perfect Laplacian on an infinite
  lattice.}
\label{plot_walldiff}
\end{figure}

The effect of the boundaries on the fixed-point Laplacian is indeed highly
local. This can be seen from the symmetry construction described above: In
the even simpler case of a single wall along the second axis
the actual operator $\rho^*(n,r)$ in terms of the translation invariant
operator $\rho^*(r)$ (i.e. the operator in absence of boundaries) is given by
\begin{equation}
\rho^*(n,r)  = \rho^*(r) - \rho^*(r_1+2n_1+1,r_2),
\end{equation}
where $n_1+1/2$ denotes the distance from the wall in lattice units
($n_1$=0,1,\dots). From the 
argument of the last term we see that the effect of the wall decreases twice as
fast with distance as the couplings of $\rho^*(r)$. The size of this
difference term in dependence of the distance from the wall is shown in
Fig.~\ref{plot_walldiff}. For $n_1 > 3$ it is below machine precision.

When truncating the operators to nearest- and next-to-nearest neighbour
couplings, one has to impose some conditions on the remaining couplings in
order to 
ensure basic properties of the operator. These conditions are the correct
dispersion relation  $E(k) = |p|$ and the limit for small momenta $
\lim_{q\rightarrow 0} \rho(q) = q^2$, leading to the sum rules
$\sum_{r_1,r_2} \rho(r_1,r_2) = 0$ and $\sum_{r_1,r_2}
(r_1^2+r_2^2)\cdot\rho(r_1,r_2) = -4$.

To check the quality of our parametrization of the
perfect Laplace operator,
we numerically solved the lattice Poisson equation $A\phi=f$ for square
boundaries with the standard
$(4,-1,-1,-1)$-Laplacian and with our parametrized Laplace operator made
up from $\rho_1$--$\rho_{6}$. For the resulting field configuration, we
determined the potential energy $E= f\sum_n \bar\phi^{(std)}_n$ 
for the field configuration $\bar\phi^{(std)}$ computed with the standard
Laplacian and $E = f_\phi\sum_n \bar\phi^{(pp)}_n + F^2 / 3 \kappa V$
for the configuration $\bar\phi^{(pp)}$ computed with the parametrized perfect
Laplacian. The additional term $F^2 / 3 \kappa V$ in the
potential energy comes from the 
constant $C$ in the RG transformation (\ref{generalrgt}) for non-zero source
\cite{hauswirth:00}. 

\begin{figure}[htb]
\begin{center}
\psfrag{yaxis}[][]{\hspace{0cm}\vspace{0.5cm}$|\Delta E|$}
\psfrag{1/n2}{\vspace{-0.3cm}$1/N^2$}
\psfrag{0.000}{\hspace{3mm}\footnotesize $0$}
\psfrag{0.005}{\hspace{-3mm}\footnotesize \ $0.005$}
\psfrag{0.010}{\hspace{-3mm}\footnotesize \ $0.010$}
\psfrag{0.015}{\hspace{-3mm}\footnotesize \ $0.015$}
\psfrag{0.020}{\hspace{-3mm}\footnotesize \ $0.020$}
\psfrag{0.00}{\hspace{2mm}\vspace{-1mm}\footnotesize $0$}
\psfrag{0.01}{\hspace{-1mm}\vspace{-1mm}\footnotesize \ $0.01$}
\psfrag{0.02}{\hspace{-1mm}\vspace{-1mm}\footnotesize \ $0.02$}
\psfrag{Standard Laplacian}{\scriptsize Standard Laplacian}
\psfrag{Parametrized Perfect Laplacian12345678}{\scriptsize Perfect
  Laplacian}  
\includegraphics[width=7cm]{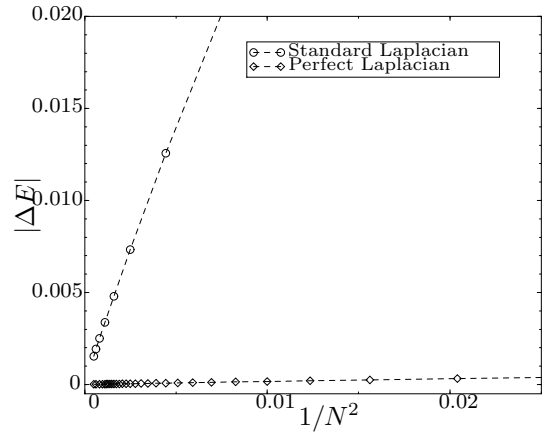}
\end{center}
\vspace{-10mm}
\caption{Relative error for the numerical solution of our
  test problem.} \label{plot_error}
\label{plot_testres1}
\end{figure}

Fig.~\ref{plot_error} shows a plot of the relative error
$\Delta E = (E-E^{(cont.)})/E^{(cont.)}$ as a 
function of the inverse lattice volume $1/V=1/N^2$. The error for our
parametrization is proportional to $1/N^2$ and is for any lattice
size smaller by a
factor of about 180 than the error of the standard Laplacian.
This factor can probably still be increased by tuning the normalization
procedure of the truncated couplings.

\section{THE POISSON EQUATION}

So far, we have worked with a constant source term. But the above construction
of a parametrized perfect Laplace operator also holds for the general Poisson
equation with a non-constant source term $J(x)$.
Consider a RG transformation of the action:
\begin{equation}
   S_J[\phi] = \frac{1}{2}\int d^dx\ [\partial_\mu\phi(x)\partial_\mu\phi(x) +
   J(x)\phi(x) ].
\end{equation}
Blocking out of continuum \cite{wilson:76,bietenholz_wiese:96} with the kernel 
\begin{equation}
\mathcal{T}_\kappa[\Phi,\phi] =
\kappa\sum_n\Bigl(\Phi_n - \int d^dx\ \omega(x-n)\phi(x)\Bigr)^2,
\end{equation}
where $\omega(x)$ is an arbitrary blocking function gives the fixed point
action, from which we read the equation of motion
\begin{equation}\label{perfpoisson}
  \sum_{n^\prime} \rho^*(n-n^\prime) \Phi_{n^\prime} = -J_n^{FP}.
\end{equation}
where the fixed point source is given by
\begin{equation}
 \tilde J^{FP}(k) = \tilde\rho^*(k) \frac{\tilde\omega(k)}{k^2} \tilde J(k).
\end{equation}
In the perfect Poisson equation (\ref{perfpoisson}) appears again the perfect
Laplacian $\rho^*(r)$ which is described by our parametrization.

\section{CONCLUSION}

The very basic examinations presented here show that the fixed-point approach
is also applicable to classical field theory problems. We show that the fixed
point Laplace operator does not care about boundaries which are a few lattice
spacings away, and therefore it is possible to give a simple parametrization
which works well in applications. Whether this approach is also fruitful for
more complicated problems remains to be investigated.

\section{ACKNOWLEDGEMENTS}
I thank Peter Hasenfratz, Ferenc Niedermayer and Urs Wenger for stimulating
discussions. This work was supported in part by Schweizerischer Nationalfonds.

\end{document}